\documentclass{ws-procs975x65}
\usepackage{graphicx}
\usepackage{hyperref}

\begin{document}

\title{Search for high-energy neutrinos from binary neutron star mergers}
\author{Nora Linn Strotjohann$^*$ for the IceCube Collaboration}

\address{DESY Zeuthen,\\
15738 Zeuthen, Germany\\
$^*$E-mail: nora.linn.strotjohann@desy.de}

\begin{abstract}
To search for transient astrophysical neutrino sources, IceCube's optical and X-ray follow-up program is triggered by two or more neutrino candidates arriving from a similar direction within 100\,s. However, the rate of such neutrino multiplets was found to be consistent with the expected background of chance coincidences, such that the data does not provide indications for the existence of short-lived transient neutrino sources. Upper limits on the neutrino flux of transient source populations are presented in Aartsen et al. (2019)\cite{minutelong2019} and we show here how these limits apply to the predicted neutrino emission from binary neutron star mergers.
\end{abstract}


\bodymatter


\section{Introduction}

The first (and so far only) detected binary neutron star merger is the object GW\,170817\cite{ligo2017} with the electromagnetic counterpart GRB\,170817A.\cite{kasliwal2017} The detected gamma-ray and multiwavelength emission indicates that a relativistic jet was launched during the merger.\cite{lazatti2018} It has been suggest that cosmic rays could be accelerated within this jet and an associated flux of high-energy neutrinos has been predicted.\cite{kimura2017, biehl2018} A search for neutrino emission from the position of GRB\,170817A was carried out by the ANTARES, IceCube and Pierre Auger observatories, but no signal was detected.\cite{icecube2017}

Here, we present limits on the neutrino emission of short-lived transients,\cite{minutelong2019} which also apply to binary neutron star mergers and are independent of the limits derived for GRB\,170817A. The results are based on data from the IceCube neutrino observatory, an ice-Cherenkov detector located at the geographic South pole.\cite{icecube2017b} An array of 5\,160 optical sensors instruments a volume of $\sim1\,\text{km}^3$ and allows to detect neutrino interactions in the energy range between $\sim100\,\text{GeV}$ and a few PeV.\cite{minutelong2019}

IceCube's optical and X-ray follow-up program (OFU program)\cite{icecube2012, icecube2015, triplet2017, minutelong2019} is taylored to search for neutrino emission in connection with gamma-ray bursts (GRBs) or core-collapse supernovae (CCSNe) with choked jets. An alert is defined as two or more tracklike events which are detected within 100\,s with a maximal angular separation of $3.5^\circ$. The OFU program currently only uses neutrino candidates from the northern sky to avoid the large background of downgoing atmospheric muons. The aim of the program is to identify astrophysical neutrino sources through optical and X-ray follow-up observations, which can be triggered within $\sim1\,\text{min}$.\citep{realtime2017, evans2015} Examples of follow-up campaigns are presented in Aarten et al. (2015)\cite{icecube2015}, Aartsen et al. (2017)\cite{triplet2017}, and Evans et al. (2015)\cite{evans2015}, but the observations so far did not identify a likely neutrino source and the number of neutrino multiplets is consistent with the expected number of chance coincidences of atmospheric neutrinos or muons.\cite{minutelong2019}

\section{Generic Limits on Short-Lived Neutrino Transients}
\label{sec:limits}

Upper limits on the neutrino flux of short-lived transients have been published in Aartsen et al. (2019)~\cite{minutelong2019} and are summarized in this section. The analysis is based on $1648.1\,\text{days}$ of IceCube data (collected between September 2011 and May 2016). Only one alert consisting of three neutrino candidates was detected during this time.\cite{triplet2017} This is consistent with the estimated background of $0.34$ chance coincidences and the resulting 90\% upper limit on the expected number of astrophysical neutrino multiplets (alerts consisting of more than two neutrino candidates) is $<4.0$.

To calculate a limit on the median neutrino luminosity of short-lived transients, an extragalactic source population is simulated and those scenarios which yield more than $4$ astrophysical neutrino multiplets are disfavored at 90\% confidence level. The source properties are chosen to be similar to long GRBs or CCSNe and several assumptions are tested to quantify their impact on the resulting limit. The assumed population properties are:

\begin{arabiclist}
 \item {\bf redshift distribution}: star formation rate for CCSN-like sources\cite{madau2014} and the distribution observed by \emph{Swift} for GRB-like sources.\cite{wanderman2010}
 \item {\bf neutrino luminosity function}: small fluctuations (one astronomical magnitude) for CCSN-like sources, observed gamma-ray luminosity function for GRB-like sources.\cite{wanderman2010}
 \item {\bf neutrino spectrum}: an $E^{-2.13}$ or $E^{-2.5}$ spectrum between $100\,\text{GeV}$ and 10\,\text{PeV} as measured for the astrophysical neutrino flux at $\sim100\,\text{TeV}$.\cite{icecube2016, icecube2015b}
 \item {\bf transient duration}: duration of long GRBs observed by the \emph{Swift} BAT.\cite{minutelong2019} 
\end{arabiclist}

\begin{figure}[h]
\begin{center}
\includegraphics[width=0.7\textwidth]{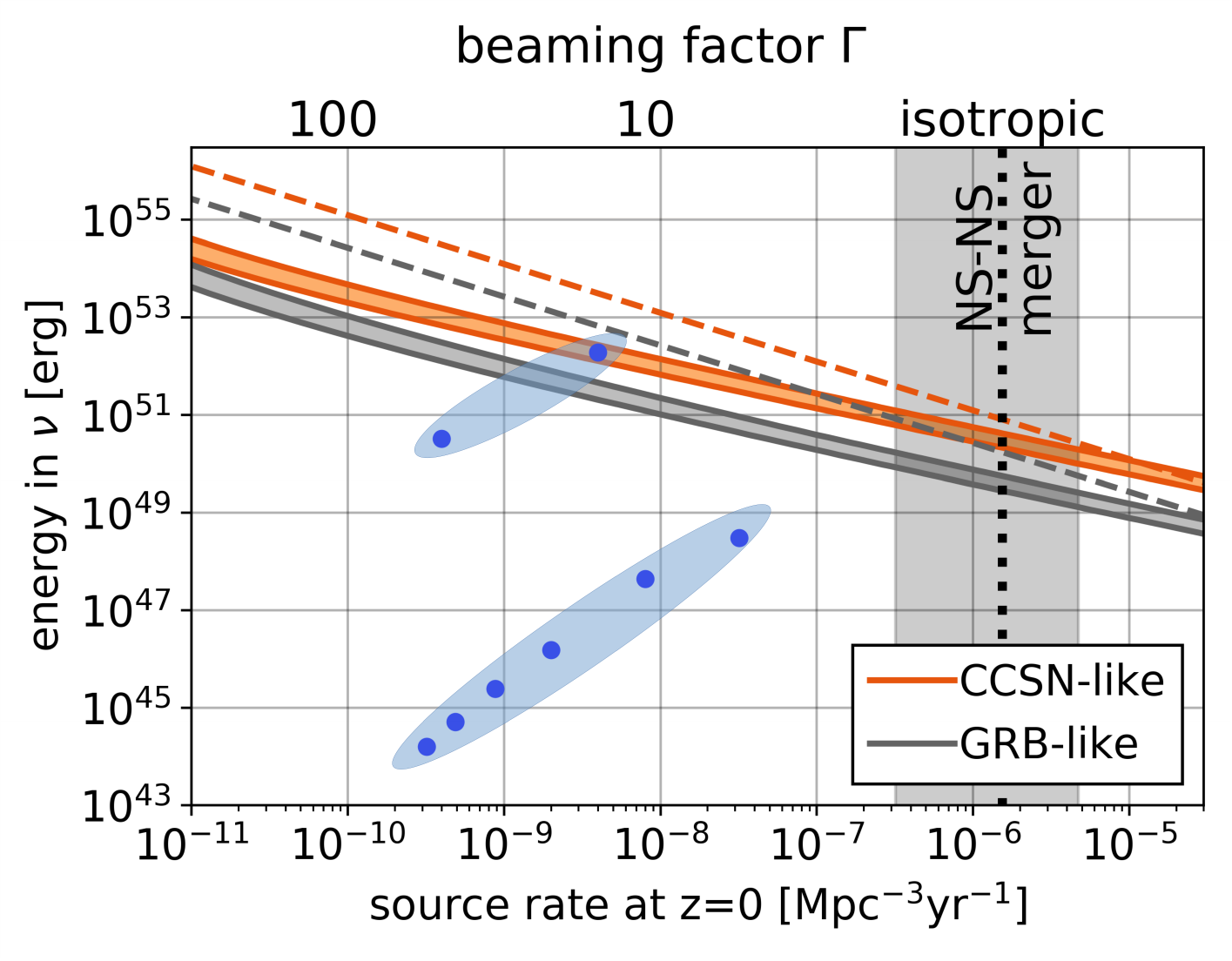}
\end{center}
\caption{Upper limits (orange and gray bands) on the median energy emitted by a transient in neutrinos (between 100\,GeV and 10\,PeV; for three neutrino flavors; compare Ref.~1). The dashed lines correspond to the source energy for which the population would produce the entire astrophysical neutrino flux for an $E^{-2.5}$ spectrum. The blue dots show the predicted high-energy neutrino emission from binary neutron star mergers, where the upper dots are for the Kimura extended emission (EE) model\cite{kimura2017} and the lower ones for the Biehl prompt flux model.\cite{biehl2018}}
\label{aba:limit}
\end{figure}

The rate density of transients and the population's neutrino flux are free parameters in the calculation. The resulting upper limits are shown as orange and gray bands in Fig.~\ref{aba:limit} and correspond to the 90\% c.l. upper limits on the energy emitted in neutrinos by the median source in the population. The limit for the GRB-like population is stronger mainly due to the steeply falling luminosity function of long GRBs.\cite{wanderman2010} The lower edge of the band shows the limit for an $E^{-2.13}$ spectrum, while the upper edge is for an $E^{-2.5}$ spectrum. The dashed diagonal lines show the source energy for which the population would produce the entire astrophysical neutrino flux assuming an $E^{-2.5}$ spectrum. The corresponding lines for an $E^{-2.13}$ spectrum are lower by a factor of 13, but are not shown here for clarity.

Rare sources, such as GRBs with a rate of $\sim4\times10^{-10}\,\text{Mpc}^{-3}\,\text{yr}^{-1}$,\cite{lien2014} can only produce a small fraction of the astrophysical neutrino flux without producing more than $4$ neutrino multiplets. This confirms the non-detections from the stacked IceCube searches for neutrinos from GRBs.\cite{grb2016, grb2017} If, on the other hand, many faint sources contribute to the flux, no neutrino multiplets are expected as even the brightest sources only yield one or two detected events. The limits are purely based on neutrino data and hence also apply to short-lived transients that commonly remain undetected, such as binary neutron star mergers or low-luminosity GRBs. They are therefore more general than stacked searches.

\section{Limits on the Neutrino Flux of Binary Neutron Star Mergers}

\begin{figure}[h]
\begin{center}
\includegraphics[width=0.7\textwidth]{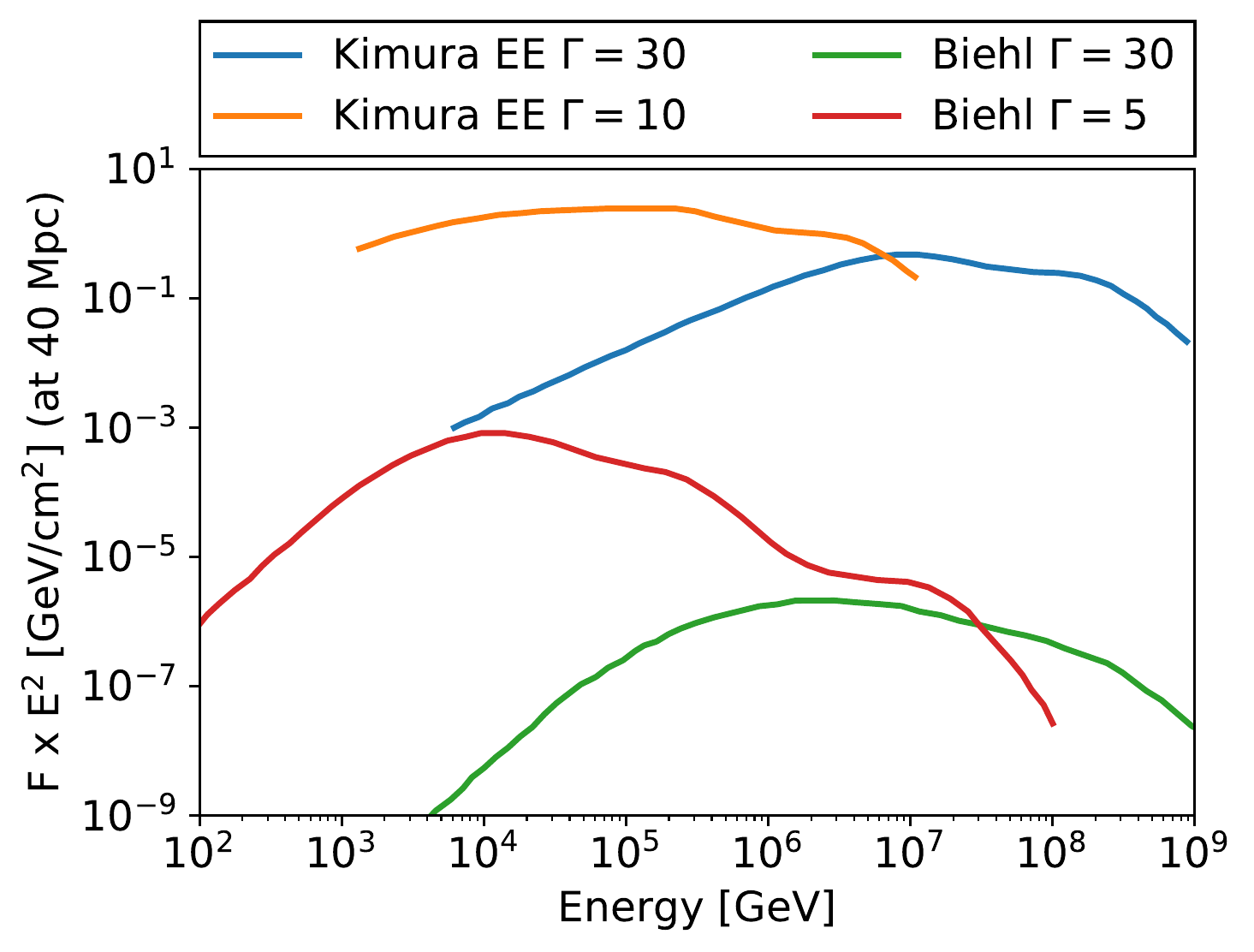}
\end{center}
\caption{Predicted neutrino spectra of binary neutron star mergers model according to the Kimura extended emission model\cite{kimura2017} and the Biehl prompt emission model\cite{biehl2018}.}
\label{aba:spec}
\end{figure}

While the results presented in Aartsen et al. (2019)\cite{minutelong2019} (and summarized in the previous section) were calculated for a generic population of neutrino sources, the limits are now compared to specific models for the high-energy neutrino emission from binary neutron star mergers. Such models have for example been published by Kimura et al. (2017)\cite{kimura2017} and Biehl et al. (2018)\cite{biehl2018}. 
There are considerable uncertainties on the model parameters, such as the beaming factor or the baryonic loading, which result in large differences between the precited neutrino emission shown in Fig.~\ref{aba:spec}. We test several models, listed in Table~\ref{tbl1}, that predict neutrino emission within 100\,s and yield a considerably higher neutrino flux between $100\,\text{GeV}$ and $10\,\text{PeV}$.

\begin{table}
\tbl{Predicted high-energy neutrino emission from binary neutron star mergers. The beaming factor $\Gamma$ determines the rate of detectable transients. The expected number of $\nu_\mu$ events (fourth column) is for the OFU event selection for sources in the northern sky. The next column lists the distance within which three or more events are expected a merger in the northern sky. The two last columns show the energy emitted in neutrinos (for the sum of all three neutrino flavors) between 100\,GeV and 10\,PeV and the effective source neutrino energy, the energy of a source with an $E^{-2.13}$ spectrum with the same expected number of OFU events.}
{\begin{tabular}{@{}lcccccc@{}}
\hline\hline
Model & $\Gamma$ & rate & $\nu_\mu$ events & max. dist. & energy in $\nu$ & eff. energy in $\nu$ \\
& & [Mpc$^{-3}$\,yr$^{-1}$]  & at 40\,Mpc & [Mpc] & [erg] & [erg] \\
\colrule
Kimura EE opt. & 10 & $4\times10^{-9}$ 	& 90 & $<220$ & $1.2\times10^{52}$ & $1.9\times10^{52}$ \\
Kimura EE mod. & 30 & $4\times10^{-10}$ 	& 1.5 & $<30$ & $8\times10^{50}$ & $3\times10^{50}$\\
Biehl optimistic & 5 	& $3\times10^{-8}$ 	& 0.02 & $<10$ & $1.6\times10^{48}$ &$3\times10^{48}$\\
Biehl moderate & 30 & $1\times10^{-9}$ 	& $1\times10^{-5}$ & $<0.1$ & $4\times10^{45}$ & $2\times10^{45}$ \\
\hline
\end{tabular}}
\label{tbl1}
\end{table}

To quantify the sensitivity of the OFU program to the predicted neutrino fluxes, the spectral shapes are multiplied with the effective area of the OFU event selection. This yields the expected number of neutrino events from a merger in the northern sky at a distance of 40\,Mpc which is given in the fourth column of Table~\ref{tbl1}. The fifth column lists within which distance three or more detected neutrinos are expected from a merger. It is neglected here that an object with fewer expected events might nevertheless produce a neutrino triplet due to Poisson fluctuations.\cite{eddington2019}

Contrary to the assumptions for the limit calculation (summarized in Sect.~\ref{sec:limits}), the spectra shown in Fig.~\ref{aba:spec} do not follow a power law. To allow a comparison an effective source energy is calculated. It corresponds to the energy of a source with an $E^{-2.13}$ spectrum between $100\,\text{GeV}$ and $10\,\text{PeV}$ for which the same number of events is expected. The effective source energy is given in the last column of Table~\ref{tbl1}. The actual source energy is calculated by integrating the spectra in Fig.~\ref{aba:spec} and is listed in the second last column of the table. The difference is less than a factor of three for all considered neutrino spectra.



The flux upper limits moreover depend on the rate density of transients at $z=0$. As shown in Fig.~\ref{aba:limit}, the limits get stronger for larger rates, because it becomes more likely that a source happens within the distance where it is detectable (listed in the fifth column of Table~\ref{tbl1}).
The rate of binary neutron star mergers was measured based on the detected event and the gray, vertical band in Fig.~\ref{aba:limit} shows the rate within its $1\sigma$ uncertainty. The considered models, however, predict collimated neutrino emission which reduces the rate of observable events by a factor of $1/(2\Gamma)$. The collimation factors and resulting rates are listed in the second and third column of Table~\ref{tbl1}. Extended gamma-ray emission is moreover only expected for every second short GRB\cite{kimura2017} (and was not detected for GRB\,170817A) which reduces the rate accordingly. Potential neutrino emission from misaligned mergers\cite{kimura2017, biehl2018} is not considered for simplicity. The redshift distribution only has a minor influence on the upper limits. When using the observed evolution of short GRBs\cite{wanderman2015} the limits become $13\%$ stronger compared to the star formation rate used for the CCSN-like population (see appendix of Aartsen et al. (2019)\cite{minutelong2019}).

The predicted rates and the source energies for binary neutron star mergers are shown as blue dots in Fig.~\ref{aba:limit}. The upper dots are for the Kimura EE model\cite{kimura2017} and the lower ones for the Biehl prompt emission model\cite{biehl2018}. The Kimura prompt emission model is not shown: It assumes a beaming factor of $1000$ and has an effective neutrino energy of $5\times10^{43}\,\text{erg}$.\cite{kimura2017} Figure~\ref{aba:limit} shows that the optimistic EE model is within reach of the OFU program. If the luminosity fluctuations between individual mergers are as large as the fluctuations observed for long GRBs in gamma rays, the lower edge of the gray band represents the relevant upper limit and the model would be disfavored at 90\% confidence limit. The models for prompt neutrino emission, however, predict fluxes and rates that are many orders of magnitude below the IceCube sensitivity.

\section{Discussion}

The observed gamma-ray emission of GRB\,170817A indicates the presence of a relativistic jet which might be able to accelerate cosmic rays and yield a neutrino flux. No neutrino emission was detected from GRB\,170817A which is expected since the jet was likely misaligned by $30^\circ$.\cite{lazatti2018} Moreover, no extended gamma-ray emission was observed from the central engine.\cite{kasliwal2017} An additional difficulty was that the source was located above the horizon for both the ANTARES and the IceCube detectors which reduces the sensitivity.\cite{icecube2017}

Independent upper limits on the neutrino emission of 100\,s-long transients are provided by IceCube's OFU program.
The sensitivity of the OFU program to a merger located at 40\,Mpc was evaluated for several models and the results are listed in Table~\ref{tbl1}. For the most optimistic model the OFU program would likely be triggered if the merger happens within $\sim220\,\text{Mpc}$ in the northern sky. To estimate how likely a nearby merger occurs during the $\sim5$ years of analyzed data, we calculate the rate of mergers which have a jet pointed at Earth. The most optimistic model predicts a relatively large rate of bright sources, such that one or several neutrino multiplets would have been detected by the OFU program.

Most prompt models, however, predict that the flux from the population of binary neutron star mergers only makes up a small fraction of the astrophysical neutrino flux. In this case, IceCube might not be sensitive enough to detect the neutrino emission from a merger. With only one detected binary neutron star merger, the uncertainties on the physical properties of these objects are still large. Models might become more precise, once more observational information is available.


Figure~\ref{aba:spec} shows that the predicted neutrino flux of several models peaks at PeV energies. Since April 2016, IceCube announces events with energies $\gtrsim100\,\text{TeV}$ to the public.\cite{realtime2017} Follow-up observations of these alerts could hence provide an alternative way to find binary neutron star mergers. The source of a single event is, however, on average located at a large distance (the median distance is $3\,\text{Gpc}$ assuming no source evolution)\cite{eddington2019}. The horizon distance of the Ligo-Virgo detector was $\sim200\,\text{Mpc}$ in run O2 and might increase to $300\,\text{Mpc}$ in run O3.\cite{ligo2017,chen2017} The optical peak magnitude of GRB\,170817A was $-15.5$ in the $R$-band.\cite{kasliwal2017} This would allow a detection within $140\,\text{Mpc}$ ($300\,\text{Mpc}$) by a telescope with a limiting magnitude of $20$ ($22$). If a single PeV neutrino is detected from a binary neutron star merger, the source is likely in most cases not detectable by gravitational wave detectors or electromagnetic follow-up telescopes.


\emph{I would like to thank Marek Kowalski, Anna Franckowiak, Vladimir Lipunov, Markus Ackermann, Daniel Biehl, Shigeo Kimura, Ignacio Taboada, and the IceCube collaboration for scientific discussions, support and inspiration.}



\begin{thebibliography}{0}

\bibitem{minutelong2019} M.~G. Aartsen et al.,
		    {\em Phys. Rev. Lett.} {\bf 122}, 051102 (2019)

\bibitem{ligo2017} B.~P. Abbott et al.,
		    {\em Phys. Rev. Lett.} {\bf 119}, 16 (2017)

\bibitem{kasliwal2017} M.~M. Kasliwal et al.,
		    {\em Science} {\bf 358}, 1559 (2017)

		    
\bibitem{lazatti2018} D. Lazzati et al.,
		    {\em Phys. Rev. Lett.} {\bf 120}, 24 (2018)


\bibitem{kimura2017} S.~S. Kimura, K. Murase, P. M{\'e}sz{\'a}ros, and K. Kiuchi,
		    {\em ApJL} {\bf 848}, L4 (2017)

		    
\bibitem{biehl2018} D. Biehl, J. Heinze and W. Winter 
		    {\em Mon. Not. Roy. Ast. Soc.} {\bf 476}, 1191 (2018)

		    
\bibitem{icecube2017} A. Albert et al.,
		    {\em ApJL} {\bf 850}, L35 (2017)

		    
\bibitem{icecube2017b} M.~G. Aartsen et al.,
		    {\em Journal of Instrumentation} {\bf 12}, P03012 (2017)

		    
\bibitem{icecube2012} R. Abbasi et al.,
		    {\em A\&A} {\bf 539}, A60 (2012)


\bibitem{icecube2015} M.~G. Aartsen et al.,
		    {\em ApJ} {\bf 811}, 52 (2015)


\bibitem{triplet2017} M.~G. Aartsen et al.,
		    {\em A\&A} {\bf 607}, A115 (2017)
		    
		    
\bibitem{realtime2017} M.~G. Aartsen et al.,
		    {\em Astroparticle Physics} {\bf 92}, 30 (2017)
		    
		    
		    
\bibitem{evans2015} P.~A. Evans et al.,
		    {\em Mon. Not. Roy. Ast. Soc.} {\bf 448}, 2210 (2015)
		    

\bibitem{madau2014} P. Madau and M. Dickinson,
		    {\em Mon. Not. Roy. Ast. Soc.} {\bf 52}, 415 (2014)
		    
\bibitem{wanderman2010} D. Wanderman and T. Piran,
		    {\em Mon. Not. Roy. Ast. Soc.} {\bf 406}, 1944 (2010)
		    		    
\bibitem{icecube2016} M.~G. Aartsen et al.,
		    {\em ApJ} {\bf 833}, 3 (2016)
		    
		    
\bibitem{icecube2015b} M.~G. Aartsen et al.,
		    {\em ApJ} {\bf 809}, 98 (2015)
		    
\bibitem{lien2014} A. Lien et al.,
		    {\em ApJ} {\bf 783}, 24 (2014)
		    

		    
\bibitem{grb2017} M.~G. Aartsen et al.,
		    {\em ApJ} {\bf 843}, 112 (2017)

\bibitem{grb2016} M.~G. Aartsen et al.,
		    {\em ApJ} {\bf 824}, 115 (2016)

\bibitem{eddington2019} N.~L. Strotjohann, M. Kowalski and A. Franckowiak,
		    {\em A\&A} {\bf 622}, L9 (2019)

\bibitem{wanderman2015} D. Wanderman and T. Piran,
		    {\em Mon. Not. Roy. Ast. Soc.} {\bf 448}, 3026 (2015)
		    
\bibitem{chen2017} H.-Y. Chen et al.,
		    {\em ArXiv e-prints} 1709.08079 (2017)

		    		    
\end{thebibliography}
\end{document}